\documentclass[12pt,a4paper]{article}
\usepackage{epsfig}
\def\eq#1{{Eq.~(\ref{#1})}}
\def\fig#1{{Fig.~\ref{#1}}}
\def\re#1{{Ref.~\cite{#1}}}

\newcommand{\beq}{\begin{equation}}
\newcommand{\eeq}{\end{equation}}
\newcommand{\beqar}[1]{\begin{eqnarray}\label{#1}}
\newcommand{\eeqar}{\end{eqnarray}}
\newcommand{\bas}{\bar{\alpha}_s}
\newcommand{\tN}{\tilde{N}}
\newcommand{\om}{\omega}
\newcommand{\f}{\varphi}
\newcommand{\C}{\mathcal{C}}
\newcommand{\s}{\mathcal{S}}
\def\arnps#1#2#3{  {\it Ann. Rev. Nucl. Part. Sci. }{\bf #1} (#2) #3}
\def\npb#1#2#3{    {\it Nucl. Phys. }{\bf B#1} (#2) #3}
 
\def\prd#1#2#3{    {\it Phys. Rev. }{\bf D#1} (#2) #3}
\def\prep#1#2#3{   {\it Phys. Rep. }{\bf #1} (#2) #3}

\def\zpc#1#2#3{    {\it Z. Phys. }{\bf C#1} (#2) #3}
\def\epjc#1#2#3{    {\it Eur.Phys.J. }{\bf C#1} (#2) #3}

\def\sjnp#1#2#3{   {\it Sov. J. Nucl. Phys. }{\bf #1} (#2) #3}

\def\jetpl#1#2#3{  {\it JETP Lett. }{\bf #1} (#2) #3}
\def\epjc#1#2#3{   {\it Eur. Phys. J.}{\bf C#1} (#2) #3}

\begin{document}
\title {{\bf Nonlinear evolution and saturation}\\
{\bf for  heavy nuclei in DIS.}}
\author{
{\bf E.~Levin\thanks{e-mail: leving@post.tau.ac.il} \quad and
\quad  K.~Tuchin\thanks{e-mail: tuchin@post.tau.ac.il}}\\[10mm]
{\it\normalsize HEP Department}\\
{\it\normalsize School of Physics and Astronomy,}\\
{\it\normalsize Raymond and Beverly Sackler Faculty of Exact Science,}\\
{\it\normalsize Tel-Aviv University, Ramat Aviv, 69978, Israel}
}
\date{January, 2001}
\maketitle 
\thispagestyle{empty}

\begin{abstract}
The nonlinear evolution equation for the scattering amplitude of colour
dipole off the heavy nucleus is solved in the double logarithmic
approximation. It is found that if the initial parton density in
a nucleus is
smaller then some critical value, then the scattering amplitude is a
function of one scaling variable inside the saturation region, whereas if
it is greater then the
critical value, then the scaling behaviour breaks down. Dependence of the
saturation scale on the number of nucleons is discussed as well.
\end{abstract}
\thispagestyle{empty}
\begin{flushright}
\vspace{-15cm}
TAUP-2664-2001\\
\today
\end{flushright}
\newpage

%%%%%%%%%%%%%%%%%%%%%%%%%%%%%%%%%%%%%%%%%%%%%%%%%%%%%%%%%%%%%%%%%%%%%%%%%%
\section{Introduction}
 
In the two previous publications\cite{LT,GEO} we solved the nonlinear
evolution equation for high parton density
QCD\cite{GLR,QIU,BALITSKY,KOV99,KOV00,MLV,CLASSIC,CL,CKB,BRAUN,KKM} in the
double
logarithmic
approximation and discussed its properties.  We showed that at high
energies there is a momentum scale $Q^2_s(y,A)$, where $y=\ln(1/x_B)$ is a
rapidity, at which the parton
evolution can no longer be described by linear evolution equation.          
This so-called saturation scale, has the following dependence on the
energy and the number of nucleons (for DIS on a nucleus): 
$Q^2_s(y,A)\sim A^\frac{2}{3} e^{4\bas y}$, which holds in the
double logarithmic approximation. In \re{LT,GEO} we derived
this result assuming that initial parton density is not very large.

The $A$ dependence of the saturation
scale can be also obtained using simple qualitative arguments. Indeed, 
as one increases energy and/or decreases virtuality of the photon,  the
density of partons inside the nucleus grows until they begin to overlap in
the transverse plane. The scale at which this occurs is proportional to
$Q^2_s(y,A)$ and can be estimated from the equation
\beq
\frac{A\pi \vec x^2 x_BG(x_B,Q^2_s)}{\pi R_A^2}=1\quad,
\eeq
where $\vec x$ is a colour dipole, $ x_BG(x_B,Q^2_s)$ is the proton
structure function and $R_A$ is the nucleus radius. Since in the
kinematical region intermediate between the linear and the saturation
ones, the anomalous dimension of the gluon structure
function $\gamma\approx\frac{1}{2}$\cite{CLASSIC,CL,BL} and $R_A\sim
A^\frac{1}{3}$,
one obtains that  $Q_s^2\sim A^\frac{2}{3}$.  

It was shown in \re{KOV00} that the initial condition for the nonlinear
evolution equation is given by the Glauber--Mueller formula\cite{GLMU}
\beq\label{INIT0}
N(\vec x,y=y_0)=1-e^{-\frac{\alpha_s\pi^2}{2N_c}\vec x^2\; 
x_BG(x_B,4/\vec x^2)\; S(b_t)}\quad,
\eeq
where $S(b_t)$ is a nucleus profile function and $b_t$ is an impact
parameter. If $b_t<R_A$, $S=\frac{3A}{2\pi R_A^2}$
(Woods-Saxon parameterization\cite{WS}), 
which implies that initially the saturation scale $Q^2_s$ is proportional
to $A^\frac{1}{3}$. So, the question arises how it becomes
proportional to
$A^\frac{2}{3}$ at large energy. 
There is an  opinion that the only dimensionful scale in the high
energy DIS is the scale set by initial condition. It would imply that the 
$A$ dependence of the saturation scale set by initial condition 
\eq{INIT0} remains the same during the
evolution since the evolution equation is conformally invariant. This
however is not the case since there is another dimensionful scale which
comes through the Glauber--Mueller initial condition
\eq{INIT0}. Indeed,  in our estimations we neglected for simplicity
the logarithmic dependence of the gluon structure function $x_BG$ of
the \emph{nucleon} on $\vec x^2$. However, this dependence on 
$\ln(\vec x^2\Lambda^2)$ introduces the second dimensionful
parameter $\Lambda$ which is
the soft QCD scale. In other words, the relevant scale for the evolution
in the nucleon is $\Lambda^2$ while that in nucleus is $A/R_A^2$.   
The later one as we see is set by the nucleus size. The former one means
that evolution can occur in a nucleon as well as in nucleus. 

The goal of our paper is to derive the saturation scale in the
semiclassical double logarithmic approximation  and discuss 
its  dependence on $A$ and energy. 
We consider the nonlinear evolution equation
derived by Yu. Kovchegov\cite{KOV99,KOV00}  in the framework of Mueller's
dipole model\cite{DIPOLE1} and solve it using  semiclassical
methods (on characteristics) in the double logarithmic
approximation. We show that there
exist a certain characteristic,  which divides the whole
kinematical  region
into two  patches: one in which the evolution is linear and another in
which the parton density saturates. The remarkable property of this
characteristic is that the dipole number remains fixed
along it. This characteristic coincides with the
critical line in the double logarithmic
approximation\cite{GLR,CLASSIC,GEOM}. Recall\cite{LT} that the critical
line is defined by $r=\left(\ln Q^2_s(y,A)/4\Lambda^2\right)$, where 
$r=\left(\ln Q^2/4\Lambda^2\right)$.

\begin{figure}
\begin{minipage}{7cm}
\begin{center}
\epsfig{file=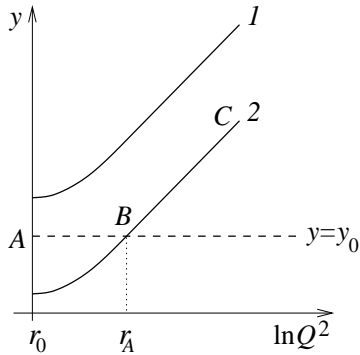, width=5cm,height=5cm}
\end{center}
\end{minipage}
\begin{minipage}{9.5cm}
\caption{{\sl The kinematical regions relevant in 
high parton density QCD. The initial condition \eq{INIT0} for the
nonlinear
evolution equation \eq{MAIN2} is set on the line $y=y_0$ (dashed line).
Solid curves 1,2 are the critical lines 
$r=\left(\ln Q_s^2/4\Lambda^2\right)$ for
different nuclei: $A_1<A_2$. Initial conditions for the solution to the
left of the critical line are set along line 1 for nucleus $A_1$ or
along line $ABC$ for nucleus $A_2$.}}\label{FIG.KINEMAT}
\end{minipage}
\end{figure}

In this paper the same strategy as in \re{LT} will be employed to solve
the evolution equation. Namely, we will solve it in the two kinematical  
regions where: (i) $\gamma$ is small ($Q^2>Q_s^2$,  $Q^2\gg\Lambda^2$) and
(ii) $\gamma$ is close to unity ($Q^2\ll Q_s^2$). Then the two solutions
will be matched in
the intermediate region. It turns out that the solution inside the
saturation region depends strongly on the initial parton density inside
nucleus. In the \fig{FIG.KINEMAT} we show two possible situations. If the
nucleus is light, then the Glauber initial condition is out of the
saturation
region. We argued in \re{LT} (and will re-derive in this paper again)  
that the parton density is constant along the critical line. This
condition sets the initial condition for the solution inside the  
saturation region. As the result, this solution is a function of only one
variable.  This case was studied in \re{LT,GEO} and will be derived
again using  another approach in Sec.~\ref{SEC.SEX}.
On the other hand, if the initial parton density is large (because the
nucleus is very heavy), then one has to match solution inside the
saturation region to both the condition on the critical line
$BC$ ($y>y_0$), and Glauber initial condition $AB$
($y=y_0$), see \fig{FIG.KINEMAT}. This cannot be done with the scaling
solution, so the scaling solution breaks down for very heavy nuclei. 
We will investigate both cases in Sec.~\ref{SEC.SEX}.

As usually we use the nonlinear evolution equation derived in \re{KOV00} 
in the framework of the dipole model\cite{DIPOLE1}. Since colour dipoles
diagonalize the scattering matrix at high energies\cite{DIPOLE,DIPOLE1}
we
consider this approach as the most clear from the physical
point of view. The evolution equation
for the dipole-nucleus forward scattering amplitude $\tN(k,y)$ in the 
momentum representation reads\footnote{
\eq{MAIN2} depends on the impact parameter $b_t$ only parametrically via
the initial condition \eq{INIT0}\cite{KOV99}. The question of $b_t$
dependence was discussed in details in \re{LT}. In this paper we address
the case $b_t=0$ which is sufficient for our arguments.}  
\beq\label{MAIN2}
\frac{\partial \tN(k,y)}{\partial y}=
\bas\hat\chi\left(\hat\gamma(k)\right)\tN(k,y)
-\bas\tN^2(k,y)\quad,
\eeq
where $\hat\chi\left(\hat\gamma(k)\right)$ is an
operator such that
\beq\label{GAMMA}
\hat\gamma(k)=1+\frac{\partial}{\partial\ln k^2}
\eeq
is an operator corresponding to the anomalous dimension of the gluon
structure function
and the operator $\hat\chi$ corresponds to the following function
\beq\label{CHI}
\chi\left(\gamma\right)=2\psi\left(1\right)-
\psi\left(1-\gamma\right)
-\psi\left(\gamma\right)
\eeq
which is an eigenvalue of the the BFKL equation. \eq{MAIN2} is
derived assuming that the typical transverse extent of the dipole
amplitude
is much smaller then the size of the target and the typical impact
parameter $x\ll R_A,b_t$. 

The amplitude in the dipole configuration space is found by employing  
the Fourier transform
\begin{eqnarray}
N(x,y)&=&\vec{x}^2\int_0^\infty dk\, k
J_0(k x)\tN(k,y)\label{TILDAN}\\
\tN(k,y)&=&\int_0^\infty \frac{dx}{x}J_0(kx)N(x,y)\quad.
\end{eqnarray}

Initial condition for this equation in the dipole configuration space
reads (see \eq{INIT0})
\beq\label{INIT}
N(x,y=y_0)=1-e^{-\exp(-r+r_A)}\quad,
\eeq
where $r_A=\ln\left(\frac{\bas\pi^2 x_BG(x_B,4/\vec x^2)S(0)}{2N_c
\Lambda^2}\right)$ and $r=\ln(Q^2/4\Lambda^2)$ for $Q^2>Q_s^2$.
In the double logarithmic
approximation one can neglect the logarithmic dependence of
$x_BG(x_B,4/\vec x^2)$ as well as $\alpha_s$ on  the dipole size, and in
addition $Q^2\approx 4/\vec x^2$. 

\eq{INIT} can be Fourier transformed to the momentum space where it reads 
\beq\label{FOURIER}
\tN(k,y)=\frac{1}{2}\Gamma_0
\left(\frac{k^2e^{-r_A}}{4\Lambda^2}\right)\quad,
\eeq
where $\Gamma_0$ is incomplete gamma function of zeroth order.

%%%%%%%%%%%%%%%%%%%%%%%%%%%%%%%%%%%%%%%%%%%%%%%%%%%%%%%%%%%%%%%%%%%%%%%%%
\section{Solution at $Q^2>Q_s^2(x_B)$.}\label{SEC.LEFT}

Function $\chi(\gamma)$ is  singular at its end-points $\gamma=0,1$. The
first one corresponds to the kinematical region $Q^2>Q_s^2(y)$,
whereas the second to the kinematical region $Q^2\ll Q_s^2(y)$.  
Expansion of $\chi(\gamma)$ near its end-points and taking the
leading contribution corresponds to the double
logarithmic approximation, i.e. to the summation of
$\alpha_s^n\ln^n(Q^2/\Lambda)\ln^n(1/x_B)$ terms to the right of the
critical line, and to the summation of
$\alpha_s^n\ln^n(Q_s^2/Q^2)\ln^n(1/x_B)$ terms to the left of it.

In the right vicinity of the $\gamma=0$ one obtains using \eq{GAMMA}
\beq
\chi=\frac{1}{\gamma}=\frac{1}{1+\partial_r}\quad.
\eeq
\eq{MAIN2}  in this approximation reads
\beq\label{RIGHT}
\tN_y(r,y)=\bas(1+\partial_r)^{-1}\tN(r,y)-\tN(r,y)^2\quad.
\eeq
In this paper we will always denote the partial derivatives by
subscripts. Note, that the following equation
\beq
(1+\partial_r)\cdot\frac{1}{k^2}\int_{\Lambda^2}^{k^2}\; dk^{'2}
\;\tN(k',y)=\tN(k,y)
\eeq
identifies the operator inverse to $1+\partial_r$. Introducing the
new function $G$ by 
\beq\label{GDEF}
G(k,y)=\int_{\Lambda^2}^{k^2}\; dk^{'2} \;\tN(k',y)\quad,
\eeq
one gets for it the following equation:
\beq\label{GEQ}
G_{ry}(r,y)=\bas G(r,y)-e^{-r}(G_r(r,y))^2\frac{1}{\Lambda^2}\quad.
\eeq
The advantage of introducing the function $G$ is that we got the nonlinear
evolution equation in which the nonlinear term has the simplest form.

Having in mind the success of the semiclassical approach to solution of
the DGLAP evolution equation at low $x_B$, we look for the solution to
the \eq{GEQ} in the form
\beq\label{SDEF}
G(r,y)=e^{S(r,y)}\Lambda^2\quad,
\eeq 
where the function $S$ is supposed to be such that
$|S_rS_y|\gg |S_{ry}|$. Equation for $S(r,y)$ is
\beq\label{SEQ}
S_r(r,y)S_y(r,y)=\bas-e^{-r+S(r,y)}S_r^2(r,y)\quad.
\eeq 
This is a partial differential equation of the first order and can be
solved by characteristics method. Denote $S_r=f$ and $S_y=\omega$. Then,
the characteristics of the \eq{SEQ} can be found by solving the system of
ordinary differential equations\cite{HILBERT}
\begin{eqnarray}
\frac{dr}{dt}&=&\om+2fe^{-r+S}\quad,\label{SYS1}\\
\frac{dy}{dt}&=&f\quad,\label{SYS2}\\
\frac{dS}{dt}&=&2f\omega+2f^2e^{-r+S}\quad,\label{SYS3}\\
\frac{df}{dt}&=&e^{-r+S}f^2(1-f)\quad,\label{SYS4}\\
\frac{d\omega}{dt}&=&-\omega e^{-r+S} f^2\quad,\label{SYS5}
\end{eqnarray}
where $t$ is the parameter along the characteristic.
Using \eq{SEQ} in \eq{SYS3} one obtains 
\beq\label{SOLS}
S=2\bas t+S_0\quad.
\eeq 
By subscript $0$ we distinguish the integration constants. It follows from
\eq{SYS4} and \eq{SYS5} that 
\beq\label{FO}
f-1=\frac{f_0-1}{\omega_0}\;\omega\equiv C\omega\quad.
\eeq
Substituting \eq{SOLS} and \eq{FO} into \eq{SYS5} results in equation
\beq\label{OMT}
\frac{d\omega}{dt}=(C\omega^2+\omega-\bas)\omega=
C\omega(\om-\om_+)(\om-\om_-)\quad,
\eeq
where 
\beq\label{OMPM}
\om_{\mp}=\frac{-1\pm\sqrt{1+4C\bas}}{2C}\quad.
\eeq

Consider some characteristic at large values of $r$ (at
$t\rightarrow\infty$). In that region
nonlinear term is small and \eq{SYS5} implies
$\om=\om_0=\mathrm{const.}$ Moreover, $f_0$ is
small and thus $\om_0=\bas/f_0$ is large.
It is straightforward to calculate $\om_+$ and $\om_-$ neglecting
nonlinear terms
\beq\label{JJ1}
\om_+=\om_0\quad \mathrm{and} \quad \om_-=\frac{\bas\om_0}{\om_0-\bas}
\quad.
\eeq
This shows that in the kinematical region of  large $r$, where nonlinear
term is small, solution of the system Eqs.~(\ref{SYS1}-\ref{SYS5}) can be  
obtained by setting $\om=\om_0=\om_+$ which yields us with the usual double
logarithmic approximation to the DGLAP and BFKL equations.  

As $t$ decreases $\om$ will no longer be equal to $\om_0$.   
Corrections stem from the nonlinear term (but $\om_0=\om_+$ still holds
as it gives correct asymptotic behaviour of solution).
Let us address the question of  what happens if the deviation  from
the linear behaviour is  small (this is equivalent to checking the
stability of the linear solution). In
other words, assume that $\om=\om_0+\delta\om$ where $\delta\om\ll\om_0$. 
Inserting this into the \eq{OMT} and integrating gives 
\beq\label{OTOM}
\om(t)=\om_0\left(1+\beta e^{-(\om_0-2\bas)t}\right)\quad,
\eeq
where $\beta$ is the integration constant, and by \eq{FO}
\beq\label{OTF}
f(t)=\frac{\bas}{\om_0}
\left(1-\frac{\om_0-\bas}{\bas}\beta e^{-(\om_0-2\bas)t}\right)\quad.
\eeq
Finally, using \eq{SEQ},\eq{OTOM} and \eq{OTF}  in \eq{SYS1} and 
\eq{SYS2}, the solution to the  system 
Eqs.~(\ref{SYS1}-\ref{SYS5}) at large $t$ reads
\begin{eqnarray}
r-r_0&=&\om_0 t-\frac{2\om_0-3\bas}{(\om_0-2\bas)\bas}\beta
\left(e^{-(\om_0-2\bas)t}-1\right)
\quad,\label{OR}\\
y-y_0&=&\frac{\bas}{\omega_0}t 
+\frac{\om-\bas}{\om_0(\om_0-2\bas)}\beta
\left(e^{-(\om_0-2\bas)t}-1\right)
\quad.\label{OY}
\end{eqnarray}

It is readily seen that if $\om_0>2\bas$ the solution for characteristic
is
stable, which means that all nonlinear effects which are introduced
through the (Glauber) initial condition die out exponentially as $t$
increases. By contrast, at $\om_0<2\bas$ the solution is not stable. 
  
This discussion reveals a remarkable property of nonlinear evolution
equation: there exist such  characteristic, that divides the whole
kinematical region into two patches\cite{LT}. One in which all nonlinear
effects are negligible, another in which saturation occurs.  
This line is given by condition $\om_0=2\bas$. Using \eq{OR} and \eq{OY}
we get the familiar equation for the critical line\cite{LT}.
\beq\label{SATSC}
r-r_0=4\bas(y-y_0)\quad.
\eeq
In the next sections we will denote this line as $r_s(y)$ or $y_s(r)$.

Excluding $\om_0$ and $t$ (at $r\rightarrow\infty$) from \eq{OR}, \eq{OY}
and \eq{SOLS} we arrive at the general solution of \eq{SEQ} in this
limit:
\beq\label{GLIM}
S(r,y)=2\sqrt{\bas(r-r_0)(y-y_0)}+S_0\quad.
\eeq
On the critical line $S|_s=r-r_0+S_0$ (we denote by symbol $|_s$ all
quantities calculated on the critical line). Therefore, by \eq{GDEF} and
\eq{SDEF} 
\beq\label{CONST}
\tN(k,y)|_s=\frac{1}{k^2}G_r(k,y)|_s=e^{S_0-r_0}=\mathrm{const.} 
\eeq
The dipole scattering amplitude is constant on the critical line.
The value of this constant is given by \eq{FOURIER}:
\beq\label{AVR}
\tN(y=y_0,r=r_A)=\frac{1}{2}\Gamma_0(1)\approx 0.1\quad,
\eeq
in accord with our expectations that the critical line is the boundary
of the linear region.

%%%%%%%%%%%%%%%%%%%%%%%%%%%%%%%%%%%%%%%%%%%%%%%%%%%%%%%%%%%%%%%%%%%%%%%%%%%
\section{Solution at $Q^2<Q_s^2(x_B)$.}\label{SEC.SEX}

In this section we  solve the nonlinear evolution equation inside the
saturation region. To this end expand $\chi$ in 
the left vicinity of the $\gamma=1$ using \eq{CHI} and \eq{GAMMA}
\beq
\chi=\frac{1}{1-\gamma}=-\frac{1}{\partial_r}\quad.
\eeq
This results in the following equation (see \eq{MAIN2})  
\beq\label{LEFT}
-\tN_{y\: r}(r,y)=\bas\tN(r,y)\left(1+2\tN_r(r,y)\right)\quad.
\eeq
Following \re{LT} we introduce the auxiliary function $\f$ by 
\beq\label{FDEF}
\tN(r,y)=\frac{1}{2}\int_r\;dr'\;\left(1-e^{-\f(r',y)}\right)\quad.
\eeq
Note, that  $r=\ln(Q^2/Q_s^2)<0$ is defined with respect to the $Q_s^2$ 
which is the largest scale in this kinematical region.
For the function $\f$ \eq{LEFT} takes especially simple form
\beq\label{EQF}
\f_{y r}=-\bas\left(1-e^{-\f(r,y)}\right)\quad.
\eeq

The idea of the solution is to split this equation into two and then to
find  such solution of one of the  equations that satisfy the other as
well. 
We are looking for the solution in the form $\f=\f(e^\s)$, where 
$\s$ is assumed to be semiclassical, i.e. $|\s_{ry}|\ll |\s_r\s_y|$.
In other words the function $\f$ is a  quickly varying function of a
slowly varying one $\s$.
The form of splitting is inspired by the solution of the previous section.
Namely, we restrict  solutions of \eq{EQF} by the requirement that 
\beq\label{SPLIT}
\s_r(r,y)\s_y(r,y)=-\C=\mathrm{const}. 
\eeq
In other words we assume that $e^\s$ satisfies the linearized equation
(\i.e.\  $\f\rightarrow 0$).

In the semiclassical approximation defined above, 
\eq{EQF} can be rewritten as
\beq\label{FS1}
\frac{d^2\f(\s)}{d\s^2}=\frac{\bas}{\C}\left(1-e^{-\f(\s)}\right)\quad.
\eeq
To solve this equation we interchange the independent $\s$ and dependent 
$\f$ variables\cite{LT}. Denote 
\beq
\zeta(\f)=\frac{d\f}{d\s}\quad.
\eeq 
Then, \eq{FS1} reads
\beq
\frac{d\zeta(\f)}{d\f}\zeta(\f)=
\frac{\bas}{\C}\left(1-e^{-\f(\s)}\right)\quad,  
\eeq
and can be easily integrated for $\zeta$ and then for $\f$. the final
result is
\beq\label{SOL1}
\s=\int^{\f(\s)}\;\frac{d\f'}
{\sqrt{\frac{2\bas}{\C}(\f'+e^{-\f'})+\frac{\zeta_0^2}{2}}}+\mathrm{const.}
\eeq

Now we turn to  \eq{SPLIT}. It is merely the DGLAP
equation in the double logarithmic semiclassical approximation.
Applying the method of characteristics 
one arrives at the following system of equations
\begin{eqnarray} 
r(t,\tau)&=&\om'_0(\tau)\, t+r'_0(\tau) \quad,\quad
y(t,\tau)=f'_0(\tau)\, t+y_0(\tau)
\quad,\quad \s(t,\tau)=-2\, t\, \C+\s_0(\tau)\quad,\nonumber\\
&&\label{SS1}\\
f'(t,\tau)&=&f_0'(\tau)=\mathrm{const}.\quad,\quad 
\om'(t,\tau)=\om_0'(\tau)=\mathrm{const}.\nonumber 
\end{eqnarray}
where we introduced the prime notation to distinguish variables of this
sections from those in the previous one. $\tau$ parameterizes the
initial condition which we specify below. By analogy with
\eq{GLIM}, \eq{SS1} give the general solution 
\beq\label{SOL2}
\s(r,y)=2\sqrt{\C}\sqrt{-(r-r_0')(y-y_0)}+\s_0\quad.
\eeq

Let us now specify the initial conditions for \eq{EQF}. They depend on the
initial parton density inside a nucleus.   If the initial density is
small (curve 1 in the \fig{FIG.KINEMAT}) then the results of
Sec.~\ref{SEC.LEFT} imply that 
\beq\label{IN1}
\f(y,r)|_s=\mathrm{const.}\quad
\eeq
The initial condition is different in the case of large initial parton
density (curve 2). In this region the initial condition reads:
\begin{eqnarray}
\f(y,r)|_s&=&\mathrm{const.}\quad,\hspace{2.8cm} \mathrm{if}\quad
               y_0<y=y_s(r)\quad,\label{IN2A}\\
\f(y,r) &=&-\ln\left(1-e^{-\exp(r-r_A)}\right)\quad,\quad
          \mathrm{if}\quad y=y_0\quad.\label{IN2B}
\end{eqnarray}
\eq{IN2B} follows from its definition \eq{FDEF} and \eq{FOURIER}.

Generally, to satisfy initial conditions one has
to find such a surface among those given by the solution \eq{SS1}
which paths through the line specified by \eq{IN1} or \eq{IN2A} and
\eq{IN2B}. To do this one should parameterize the
initial condition using parameter $\tau$. Then calculate $f_0'(\tau)$ and 
$\om_0'(\tau)$ using  the following equations\cite{HILBERT}:
\beq\label{EQ1}
\frac{d\s_0}{d\tau}=f_0'\;\frac{dr_0'}{d\tau}+
\om_0'\;\frac{dy_0'}{d\tau}\quad,
\eeq
and 
\beq\label{EQ2}
f_0'\;\om_0'=-\C\quad.
\eeq
 The desired solution of the Cauchy problem
$\s(y,r)$ is then obtained by excluding $t$ and $\tau$ from \eq{SS1}. 

%%%%%
\eq{EQ1} and \eq{EQ2} can be easily solved numerically, but to give a
transparent analytic treatment we notice that the nonlinear corrections in
the kinematical 
region $Q^2<Q^2_s$ are so large that $e^{-\f}\ll 1$. Indeed, by \eq{IN2A}
and \eq{IN2B}, the value of $\f$ on the critical line $r=r_s(y)$ is
\beq\label{F0}
\f_0=-\ln\left(1-e^{-1}\right)\approx 0.5\quad,
\eeq 
which follows from the requirement that the point $(r_A,y_0)$ lies on the
saturation scale (see \fig{FIG.KINEMAT}). 
Then \eq{EQF} can be easily integrated
\beq\label{MODSOL}
\f(y,r)=-\bas y\, r+f_1(y)+f_2(r)\quad,
\eeq
where $f_1$ and $f_2$ are arbitrary functions. 

Consider two different initial conditions.
\begin{enumerate}
\item\noindent
In the case of large initial parton density the initial condition
Eqs.~(\ref{IN2A},\ref{IN2B}) read
\begin{eqnarray}
\f(r,y=y_0)&=&-\bas y_0\, r+f_1(y_0)+f_2(r)=
-\ln\left(1-e^{-\exp{(r-r_A)}}\right)\quad,\\
\f(r,y=y_s(r))&=&-\bas y_s(r)\, r+f_1(y_s(r))+f_2(r)=\f_0\quad,
\end{eqnarray}
and is satisfied by the following function
\begin{eqnarray}
\f(y,r)&=&-\bas (y-y_0)\, r+\bas (y-y_0)\, r_s(y)-
\ln\left(1-e^{-\exp{(r-r_A)}}\right)+\nonumber\\
&&
\ln\left(1-e^{-\exp{(r_s(y)-r_A)}}\right)+\f_0\quad.
\label{MMM}
\end{eqnarray}  

\item\noindent
If the initial parton density is small, then in addition to the initial
condition \eq{IN1} for the function $\f$, we have to specify the initial
condition for its derivative. Note, that to the right of the critical
line $\f=2\sqrt{\bas(y-y_0)r}-r+\f_0$. Thus $\f_r|_s=-\frac{1}{2}$. This
yields 
\begin{eqnarray}
\f(r,y=y_s(r))&=&-\bas\, y_s(r)\, r+f_1(y_s(r))+f_2(r)=\f_0\quad,\\
-\frac{1}{2}&=&-\bas y_s(r)+f_{2r}(r)\quad.
\end{eqnarray}
The solution is 
\beq\label{SCALSOL}
\f(r,y)=\frac{1}{8}(r-4\bas y)^2-\frac{1}{2}(r-4\bas y)+ \f_0
=\frac{1}{8}z^2-\frac{1}{2}z+\f_0\quad.
\eeq
It coincides with the asymptotic solution obtained in \re{LT}.
\end{enumerate}

We see how the introduction of the auxiliary function $\f$ 
simplifies the solution of \eq{LEFT}. The main point is that while the
value  of the amplitude $\tN$ on the critical line\eq{AVR} is intermediate
between perturbative and saturation regions and thus all terms in \eq{LEFT}
are of  the same order, the value of $\f$ on the same line \eq{F0} is such
that \eq{EQF}
can be greatly simplified: all nonlinear corrections to $\tN$ are
described with good approximation by just a constant in the equation for
$\f$.

%%%%%%%%%%%%%%%%%%%%%%%%%%%%%%%%%%%%%%%%%%%%%%%%%%%%%%%%%%%%%%%%%%%%%%%%%
\section{Discussion}

In Sec.~2 we found that in the kinematical region to the right of the
critical line  the nonlinear corrections are small
in the double logarithmic approximation. This result is in agreement with
\re{LT}. The evolution equation was solved in the semiclassical
approach using the method of characteristics. The characteristics
intersect the line  $y=y_0$ with different angles, the tangents of which
are given by the values of $f_0$. At low $f_0$ (large $\om_0$)
characteristics run in the linear kinematical region. As  $\om_0$
decreases characteristics approach the critical line and finally, at      
$\om=2\bas$ there exists the characteristic which runs along the boundary
of the saturation region. In the double logarithmic approximation
the  characteristic with $\om_0=2\bas$ is merely the
critical line. On this critical line the scattering amplitude was
found to be constant (see \eq{CONST}). 

To find the $A$-dependence of the saturation scale we have to require
that the critical line (which is the characteristic at large
$r$) satisfy
the following condition
\beq\label{GLSC}
r_s(y=y_0)=r_A\quad,
\eeq
which means that the Glauber initial condition \eq{INIT} itself provides a
saturation scale ($r_A$). However, when $r_s\sim r_A$ the value of $\om_0$
may differ from that at $r\rightarrow\infty$ (i.e.\ $2\bas$).
This is because our approximation that $|S_rS_y|\gg |S_{ry}|$ which led to
\eq{SEQ} breaks down at $r_s\approx r_A$.
The exact solution of the (linearized) evolution equation  
which satisfies the (linearized) initial condition (see \eq{INIT}) is
\beq\label{EXS}
\tN(r,y)=\int\; \frac{df}{(2\pi i)f}e^{-r+r_A+\bas (y-y_0)\frac{1}{f}+rf}
\stackrel{\mathrm{DLA}}{\approx} e^{2\sqrt{\bas r(y-y_0)}\,+r_A-r}
\eeq
(the integartion contour runs along the line parallel to the imaginary
axes to the right of all singularities of the integrand). 
Assuming that the remarkable property $\tN|_s=\mathrm{const.}$ holds up
to $r=r_A$ one gets (neglecting logarithmic contributions)
\beq
2\sqrt{\bas r(y-y_0)}+r_A-r=\phi=\mathrm{const.}
\eeq
Thus,
\beq\label{SATW}
\bas (y_s(r)-y_0)=\frac{(r-r_A)^2}{4r}\quad,
\eeq
where $\phi=0$ to satisfy $r_s(y_0)=r_A$.
At $r\gg r_A$ the $A$ dependance of the saturation scale is given by
\beq
r=4\bas y+2r_A\quad, 
\eeq
in full correspondance with our previous work\cite{LT} and esimations in
the Introduction.

In Sec.~3 we showed that the initial parton density  in the
nucleus plays very important role in the saturation
regeme dynamics. Namely, if $A$ is such that initial parton density is
smaller then some
critical value, then the solution inside the saturation region depends only on
one variable $Q^2/Q_s^2(x_B)$. This scaling behaviour was predicted by
many authors\cite{LT,MLV,BL,GEOM} and observed in experimental
data\cite{GEOM}. It turns out that if the initial parton density is larger
then the critical value, the scaling behaviour breaks down. Let us estimate
at what $A$ the parton density reaches its critical value. It is seen in
\fig{FIG.KINEMAT} that this occurs when $r_A=r_0$, where $r_0$ defines
the initial point of evolution, i.e.\ it satisfies $\alpha_s(r_0)\ll 1$.
Using definition of $r_A$ we get
\beq
\frac{1}{2}\sigma_\mathrm{dipole}S(0)=1\quad,
\eeq
where $\sigma_\mathrm{dipole}$ is a cross section for the
small-dipole--nucleon interaction. Since at $x_B=10^{-2}$ (i.e.\ $y=y_0$) 
$\sigma_\mathrm{dipole}\approx\frac{1}{2}\sigma_{\pi p}$\cite{Nucleon} 
and $S(0)=2R_A\,\rho$, where $\rho$ is a nuclear density, we have
\beq
\frac{1}{2}\sigma_{\pi p}\,\rho R_A=1\quad,
\eeq 
Taking $\sigma_{\pi p}=24$~mb, 
$\rho=0.17$~fm$^{-3}$ and $R_A=1.1\, A^{1/3}$~fm one gets
$A_\mathrm{crit.}\sim 70$. The direct estimation using double logarithmic
formulae are not reliable since they are valid up to some logarithmic
corrections and perhaps a numerical factor, which are quite important
since $A$ enters to the $1/3$ power. 

Concluding this paper, we would like to emphasize that we found that the
saturation scale for DIS on nuclei lighter then $A\sim 70$ scales like
$Q_s^2\sim A^{2/3}$ at all energies, whereas for DIS on nuclei heavier
then $A\sim 70$ it scales like $Q_s^2\sim A^{1/3}$ when the evolution
starts and then reaches the same scaling behaviour as for light nuclei 
$Q_s^2\sim A^{2/3}$ at very high energy. 

%%%%%%%%%%%%%%%%%%%%%%%%%%%%%%%%%%%%%%%%%%%%%%%%%%%%%%%%%%%%%%%%%%%%%%%%%%%%
\newpage

\vskip0.3cm
{\large\bf Acknowledgements}
\vskip0.3cm
We wish to  acknowledge the interesting and fruitful discussions with
E. Gotsman, U. Maor and L. McLerran. Our special thanks go to
Yu. Kovchegov for raising the questions
discussed in this paper and for stimulating discussions.
This research was supported in part by the Israel Academy of Science and
Humanities and by BSF grant \#98000276.
%%%%%%%%%%%%%%%%%%%%%%%%%%%%%%%%%%%%%%%%%%%%%%%%%%%%%%%%%%%%%%%%%%%%%

\end{document}